\documentclass[preprint,12pt]{elsarticle}

\usepackage{amssymb}
\usepackage{amsmath}
\usepackage{cleveref} 
\usepackage{siunitx} 

\usepackage{xcolor} 

\journal{Nuclear Physics B}

\begin{document}

\begin{frontmatter}



\title{Development of a nonlinear plasma lens \\ for achromatic beam transport}


\author[1]{P. Drobniak\corref{cor1}} 
\ead{pierre.drobniak@fys.uio.no}
\cortext[cor1]{Corresponding author}

\author[1]{E. Adli} 
\author[1]{H. Bergravf Anderson} 
\author[2]{A. Dyson} 
\author[3]{S. M. Mewes} 
\author[1]{K. N. Sjobak} 
\author[3]{M. Th{\'e}venet} 
\author[1]{C. A. Lindstr{\o}m} 

\affiliation[1]{organization={Department of Physics, University of Oslo},
            postcode={0316}, 
            state={Oslo},
            country={Norway}}

\affiliation[2]{organization={Department of Physics, University of Oxford},
            addressline={Clarendon Laboratory, Parks Road},
            postcode={OX1 3PU},
            city={Oxford},
            country={United Kingdom}}
            
\affiliation[3]{organization={Deutsches Elektronen-Synchrotron DESY},
            addressline={Notkestr. 85}, 
            postcode={22607}, 
            city={Hamburg},
            country={Germany}}

\begin{abstract}
We introduce the new idea of a nonlinear active plasma lens, as part of a larger transport lattice for achromatic electron beam transport. The proposed implementation is based on using the Hall effect in a plasma and is motivated by 1D-hydrodynamic simulations. The manufactured design is presented, including its undergoing experimental characterisation on the CLEAR beam-line at CERN.
\end{abstract}




\end{frontmatter}



\section{Introduction}

Plasma acceleration \cite{Veksler1956,Tajima1979,Chen1985,Ruth1985} is a promising technology for next-generation electron accelerators, mainly due to the high gradient they can sustain, orders of magnitude above conventional radio-frequency (RF) cavities. Currently, plasma accelerators typically offer beams with energy ranging from the MeV to a few GeV-level \cite{gonsalves2019petawatt}, with a few-percent energy spread, from pC to nC of charge, at repetition rate around $10\,$Hz and duration of a few fs. Such beams may already be promising candidates for applications including radiation therapy with electrons \cite{labate2020toward} and x-ray free-electron lasers \cite{espinos2021high}. In order to reach much higher beam-energy gains than the  \SI{\sim10}{GeV} available from a single plasma stage, such as the 20--\SI{100}{GeV} required for future strong-field quantum electrodynamics (SFQED) experiments \cite{Abramowicz2021} or the TeV-level energies required for particle colliders \cite{Shiltsev2021,Foster2023}, multiple stages must be used \cite{Lindstrom2021}.

In conventional RF accelerators, a series of quadrupoles is used to keep the beam focused. However, using the same technology to stage plasma accelerators is not suitable due to beam chromaticity. One compact alternative is to use so-called \emph{active plasma lenses} \cite{panofsky1950,van2015active}. These devices are composed of a volume, filled up with gas, where a longitudinal discharge in the kA-level creates an azimuthal B-field distribution focusing in both planes simultaneously.

A first experimental staging of plasma accelerators was achieved by Steinke \textit{et al.} in 2016 \cite{steinke2016multistage}, where two laser-plasma accelerators (one gas-jet injector for the electron source and one discharge capillary for an additional acceleration) were coupled by an active plasma lens. This result represents a key milestone for plasma staging. However, the coupling efficiency was poor, with a few percent of the charge remaining at the end of the line. One origin for this is the \emph{chromatic} behaviour of the lens \cite{Migliorati2013}, focusing each beam energy at different focal lengths.

\begin{figure}[h]
\centering
\includegraphics[width=1\linewidth]{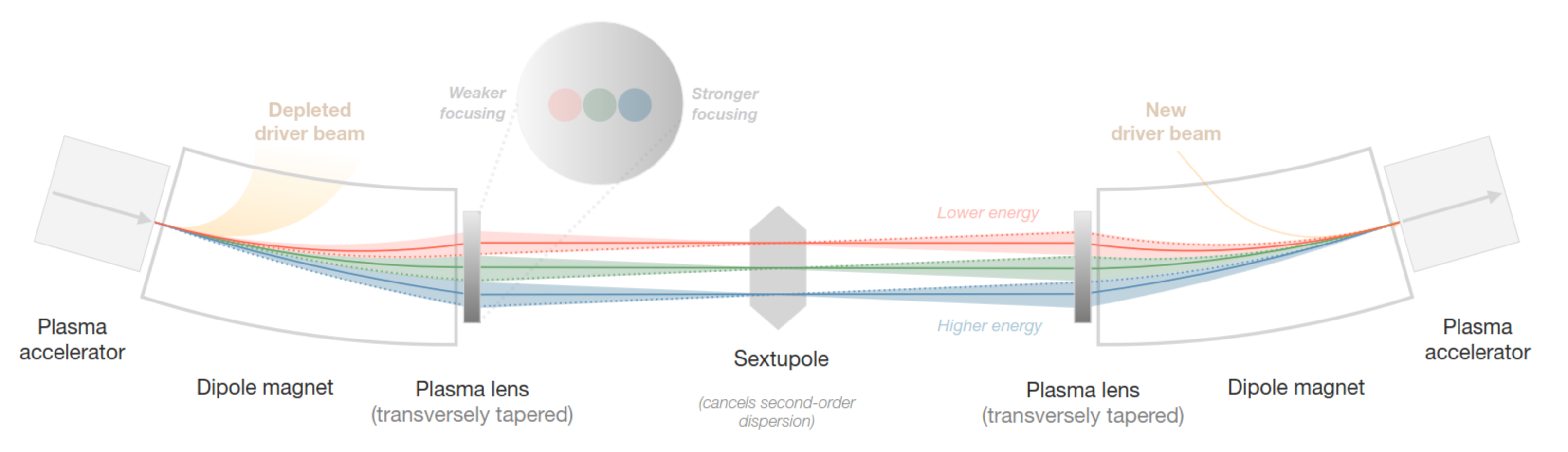}
\caption{Schematic layout of an achromatic lattice using dipole magnets to disperse the beam onto a nonlinear plasma lens. Here, the focusing strength varies transversely, and is tuned such that the focal length is equal for all energies. A mirror symmetric setup ensures that both nonlinear forces as well as the dispersion is canceled. The central sextupole is used to cancel the second-order dispersion. From Ref.~\cite{LindstromSparta}.}\label{fig:lattice}
\end{figure}

To solve this issue, an \emph{achromatic} lattice has been proposed \cite{LindstromSparta}, as described in Fig.~\ref{fig:lattice}. It requires two nonlinear plasma lenses, described in more detail below.

This article describes the planning and preliminary results of the development, both theoretical and experimental, of a nonlinear plasma lens -- a key part of the ERC project SPARTA \cite{spartaProject2023}. Section~\ref{sec:nonlinear} introduces the general idea of a nonlinear plasma lens, useful for the achromatic lattice proposed in Fig.~\ref{fig:lattice} together with potential solutions to realise it. Hydrodynamic simulations are then presented (Sec.~\ref{sec:simulation}), which motivate the idea proposed in Sec.~\ref{sec:nonlinear}. The design we developed is described (Sec.~\ref{sec:design}) and the external B-field distribution required by the design is simulated and measured (Sec.~\ref{sec:external}). Lastly, the lens characterisation is presented (Sec.~\ref{sec:setup}), as well as an outlook towards the next steps of the project (Sec.~\ref{sec:conclusion}).

\section{The nonlinear plasma lens}
\label{sec:nonlinear}

\begin{figure}[h]
\centering
\includegraphics[width=0.9\linewidth]{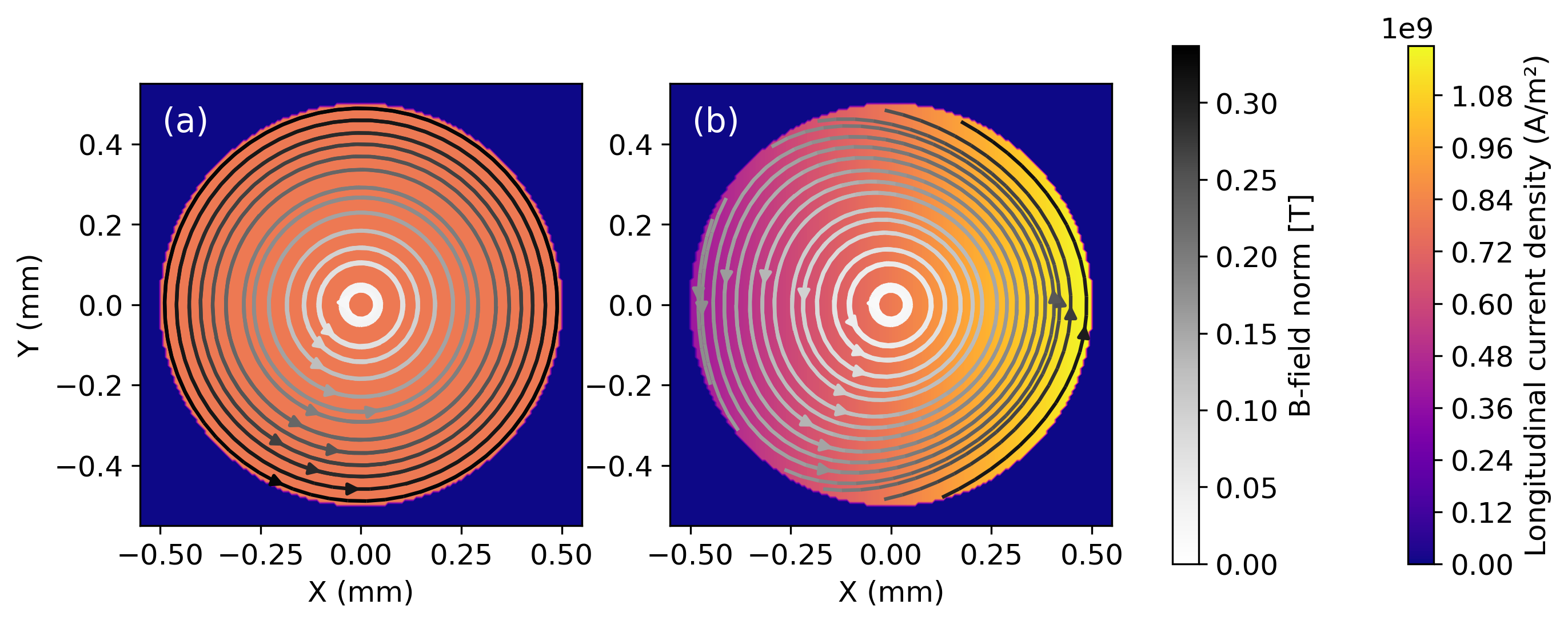}
\caption{Theoretical B-field distribution in the linear case (a) and nonlinear non-radially-symmetric case (b). A circle with radius \SI{500}{\micro\m} is added to indicate the capillary walls. Coefficients for the plots are: $g_0 = \SI{500}{T/m}$, $D_x = \SI{1}{mm}$ (note that this is exaggerated by a factor 10 or so for better visibility of the nonlinearity). }\label{fig:egg}
\end{figure}

An ``ideal" plasma lens, known as a \emph{linear} plasma lens, has a longitudinal current density that is transversely uniform $\overrightarrow{j_0} = j_{0,z} \overrightarrow{e_z}$ within the circular aperture of the lens. Using Ampère-Maxwell's law $\nabla \vec{B} = \mu{}_0 \overrightarrow{j}$ (with $\vec{j}$ being the current density), we get a purely azimuthal field [see Fig.~\ref{fig:egg}(a)], linearly growing with the distance from axis:
\begin{align}
B_x^{\mathrm{lin}} = -g_0 y \label{eq:B_lin_x}\\
B_y^{\mathrm{lin}} = +g_0 x \label{eq:B_lin_y}
\end{align}
with $g_0 = \mu{}_0 \|\overrightarrow{j_{0}}\|/2$ the focusing gradient for this case.

However, the current density is not always uniform. Cooling at the lens walls leads to a hotter core with a higher conductivity ($j \propto E\sigma \propto T^{3/2}$ \cite{Bobrova2001,van2017nonuniform}, with $E$ the electric field, $\sigma$ the electrical conductivity and $T$ the temperature), so a nonuniform distribution of the current density (validated experimentally \cite{Rockemann2018,lindstrom2018emittance}). This causes the lens to be nonlinear (nonlinear focusing strength), but still with an azimuthal symmetry.


While the above indicates the possibility of nonlinear plasma lenses, it is not a suitable nonlinearity for the achromatic lattice in Fig.~\ref{fig:lattice}. The motivation of this article is to propose a nonlinear lens with focusing strength varying in \emph{one} direction ($x$) and constant in the other direction ($y$) (see Fig. \ref{fig:lattice}). Let us consider a beam, dispersed in $x$ by a dipole (Fig.~\ref{fig:lattice}) with dispersion $D_x$. The centroid of each energy slice of the beam thus enters the lens with a different $x$-position. In order to provide the same focal length to each energy component, the focusing gradient must be $g(x,y) = g_0 (1+x/D_x)$. The distribution satisfying such a condition is:
\begin{align}
B_x^{\mathrm{nonlin}} &= -g_0 \left( y + \frac{1}{D_x} xy \right) \label{eq:B_nonlin_x} \\
B_y^{\mathrm{nonlin}} &= +g_0 \left( x + \frac{1}{D_x}\frac{x^2 + y^2}{2} \right) \label{eq:B_nonlin_y}.
\end{align}
Typical values for $g_0$ are in the 100--\SI{1000}{T/m} range and we want $1/D_x$ values around 1--10\% per lens radius (i.e., around \SI{\sim 100}{m} for a typical \SI{1}{mm}-diameter lens). The shape of the corresponding field is displayed in Fig.~\ref{fig:egg}, where the current is added in the background. The plasma lens described here (non-azimuthally symmetric) will from now on be referred to as \emph{nonlinear}.

\begin{figure}[h]
\centering
\includegraphics[width=1.0\linewidth]{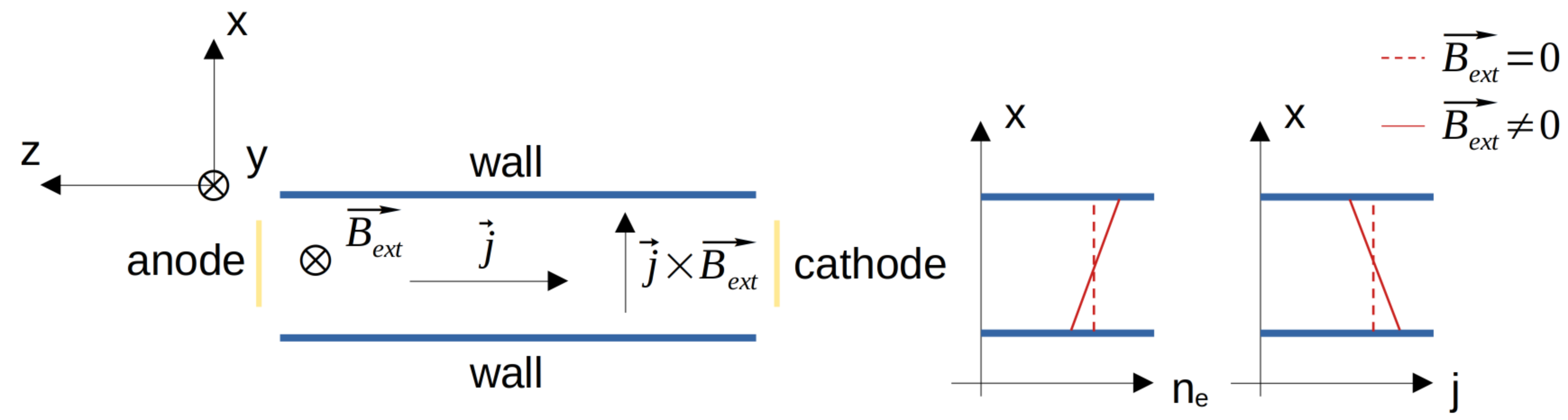}
\caption{Principles of the Hall effect in a plasma. The external B-field $\overrightarrow{B}_{\mathrm{ext}}$ acts on the current $\vec{j}$ through the Lorentz force $\vec{j} \times \overrightarrow{B}_{\mathrm{ext}}$. The theoretical effect on the electron density $n_e$ and $j$ is added, assuming that $j \propto 1/n_e$. Based on Kunkel (1981) \cite{kunkel1981hall}.}\label{fig:hall}
\end{figure}

Our proposed solution to obtain this B-field distribution is to make use of the Hall effect, inspired by Kunkel \cite{kunkel1981hall}. An external B-field $\overrightarrow{B}_{ext}$ is applied to a plasma which reacts by re-arranging its electron density $n_e$ transversely (Hall effect), as presented in Fig.~\ref{fig:hall}.

Since the current density $\vec{j}$ depends on the temperature $T_e$, which again depends on the electron density $n_e$, if one acts on $n_e$, one may produce a nonuniform current density. Optimally, a current density which is linearly dependent on $x$ is required to satisfy \cref{eq:B_nonlin_x,eq:B_nonlin_y}: 
\begin{equation}
j_z^{\mathrm{nonuni}} = \frac{2g_0}{\mu_0}(1+\frac{1}{D_x}x).
\label{eq:j_nonunif}
\end{equation}

\section{Simulation}
\label{sec:simulation}

Hydrodynamic simulations were performed in collaboration with DESY, using a modified version of the COMSOL simulation framework presented in \cite{mewes2023demonstration}. It models the ionisation of a gas (currently only $H_2$) from a time-varying input current. The plasma is modeled as a single fluid, in a non-equilibrium reaction system (ionisation and recombination modeled), with two temperatures, one for the electrons and one for the ``heavies" (ions and neutrals). Working with a 1D-Cartesian model assumes slab geometry, i.e., that the capillary is infinitely long and infinitely wide (in $y$ for instance). As of yet, only preliminary 1D-simulations (in $x$) have been performed (for computation cost reasons), where we added the external B-field (in $y$) in search for a Hall effect in the plasma.

\begin{figure}[h]
\centering
\includegraphics[width=1.0\linewidth]{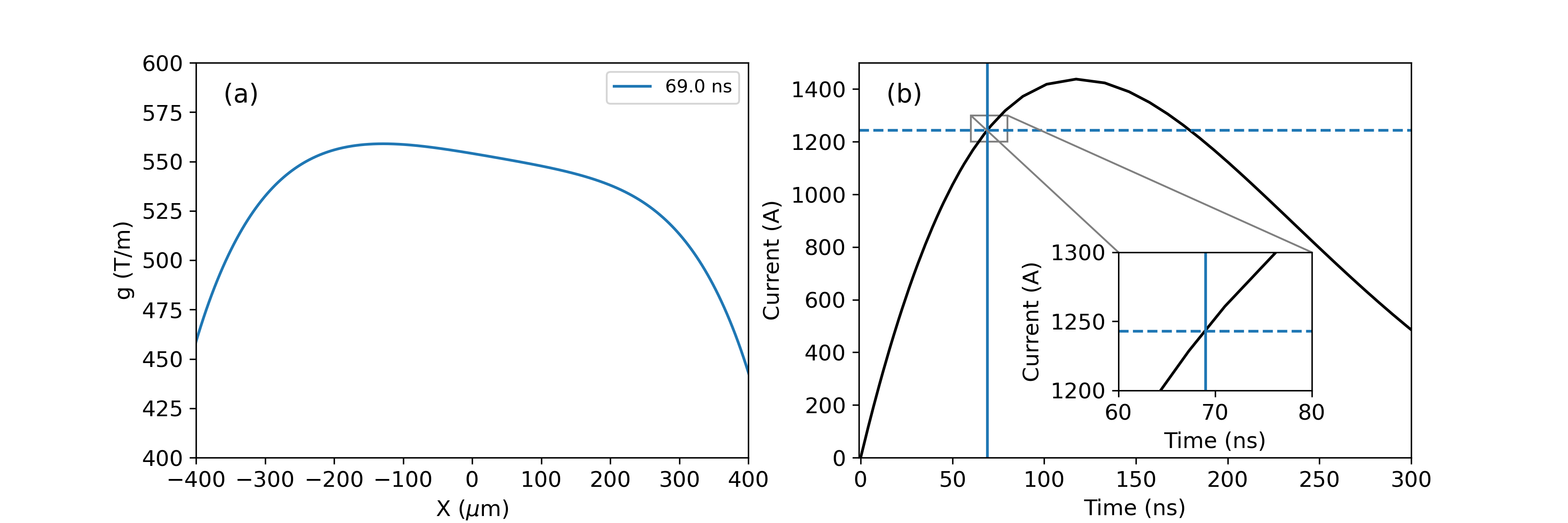}
\caption{COMSOL hydrodynamics simulation results in 1D displaying: (a) the focusing strength $g = dB_y / dx$ and (b) the current profile used as input in the simulation. The species used is $H_2$, at \SI{13}{mbar}, and is fully ionised at the time-step presented in (a). The capillary is $1\,$mm wide. The external B-field is along $y$ and set to \SI{10}{mT}.}\label{fig:MHD}
\end{figure}

First simulation results are presented in Fig.~\ref{fig:MHD}.
Within a transverse window of $[-150,+150]$~\SI{}{\micro\m}, the value of $g$ lies around \SI{550}{T/m}, with an $x$-derivative corresponding to a linear term matching a $D_x$ of \SI{10}{mm}, which is what we wish to achieve.


Outside of the $[-150,+150]$~\SI{}{\micro\m} window, the distribution of $g$ starts dropping, which originates from the fast thermal dynamics of $H_2$ (already observed experimentally without external B-field \cite{lindstrom2018emittance}), transversely rearranging close to a thermal steady-state. We think this could be mitigated by using a heavier gas. A model for this is currently under development, which is more complex than for hydrogen since a heavier gas has more ionization states.


\section{Hardware design}
\label{sec:design}

\begin{figure}[h]
\centering
\includegraphics[width=0.7\linewidth]{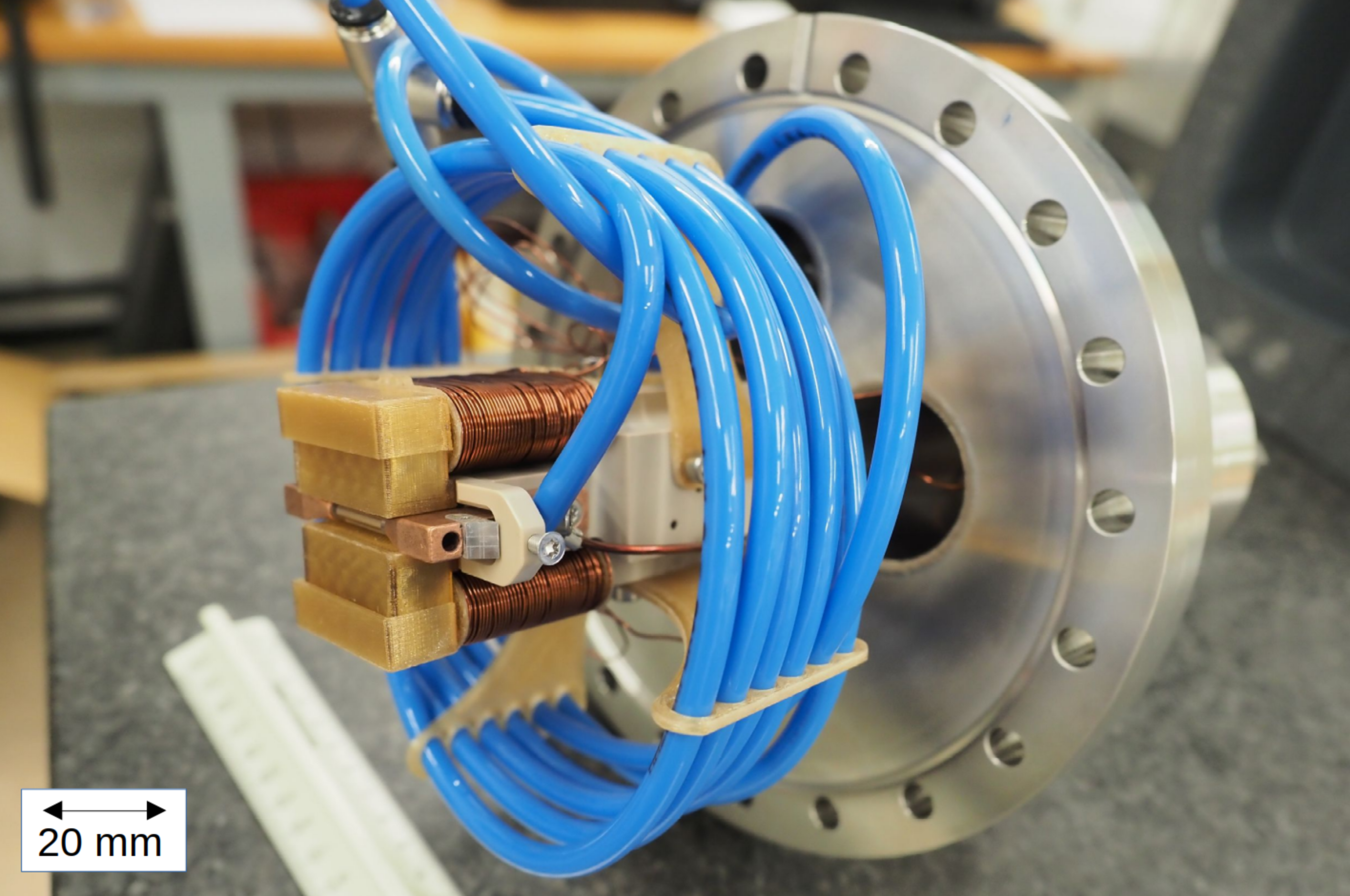}
\caption{Plasma-lens design developed at the University of Oslo and prepared for integration in the CLEAR beam-line. The capillary is held between two electrodes (copper) and an external electromagnet creates the external B-field for generating the Hall effect in the plasma. Gas is injected on each side of the capillary.}\label{fig:design}
\end{figure}

A new plasma-lens design is presented in Fig.~\ref{fig:design}. This was developed and manufactured at the University of Oslo, in collaboration with the Instrumentation Laboratory (I-Lab).

It is composed of a capillary, held between two electrodes (a tubular capillary setup inspired by recent designs from DESY~\cite{wood:ipac2024-mopr40}). Gas is injected on both sides, and a vertically aligned external B-field is generated by an electromagnet surrounding the capillary. A small mirror (OTR screen) is attached close to the entrance electrode for alignment and beam-size measurements. In order to avoid sparking between the conducting surfaces of the electrodes and the magnet poles (only a few mm apart), we covered the magnet poles with 3D-printed insulating material. The entire assembly is supported by a PEEK holder, for integration to the flange of a vacuum chamber. For compactness and easy mounting, the gas pipes (approx.~\SI{1}{m} each) that feed the capillary are supported by 3D-printed pipe holders.


\section{External B-field distribution}
\label{sec:external}

In order to assess the validity of our electromagnet design, we performed magnetostatic simulations with COMSOL \cite{comsol2021}, with the aim to ensure a uniform and vertical (along $y$) external B-field distribution. The model is built by assuming a uniform magnetisation of each arm (upper and lower) of the yoke.

\begin{figure}[h]
\centering
\includegraphics[width=1.0\linewidth]{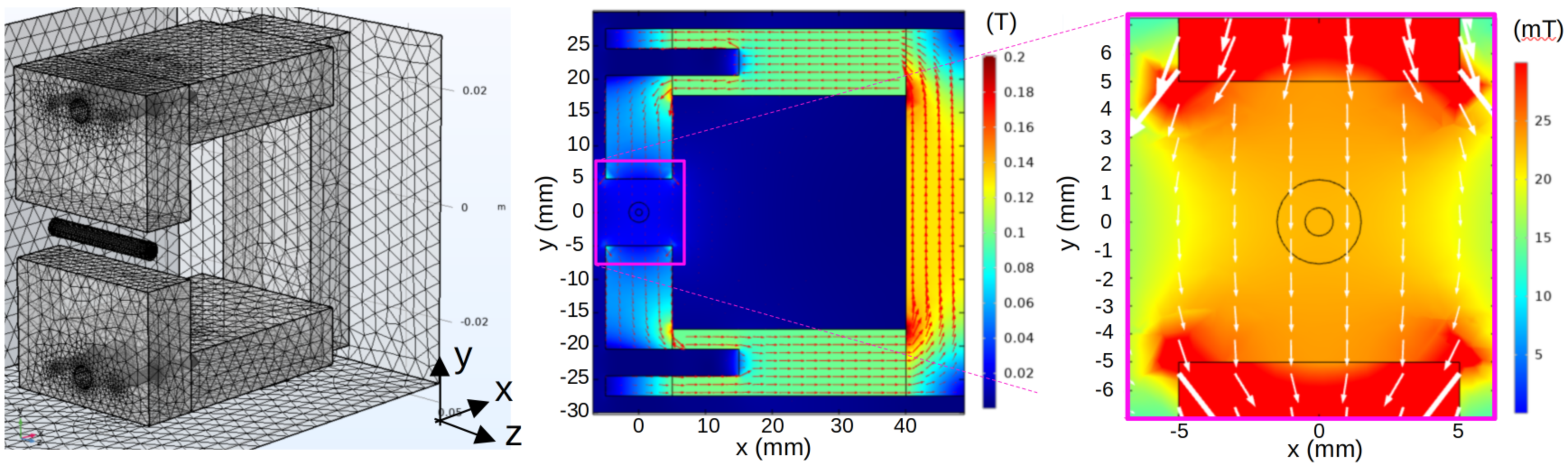}
\caption{COMSOL \cite{comsol2021} magnetostatic simulation results of the electromagnet design.}\label{fig:electromagnet_comsol}
\end{figure}

Simulation results for a B-field of \SI{100}{mT} inside the yoke (theoretically achievable with a small current in the coils and low permeability material) in each arm are presented in Fig.~\ref{fig:electromagnet_comsol}.

At the location of the capillary, the external B-field remains vertical and its value is constant $22.8 \pm \SI{0.1}{mT}$. Compared to the \SI{10}{mT} from our hydrodynamic simulations, this value is satisfactory. The transfer from yoke arm to capillary is from \SI{100}{mT} to \SI{22.8}{mT}, so roughly 23\%.

\begin{figure}[h]
\centering
\includegraphics[width=1.0\linewidth]{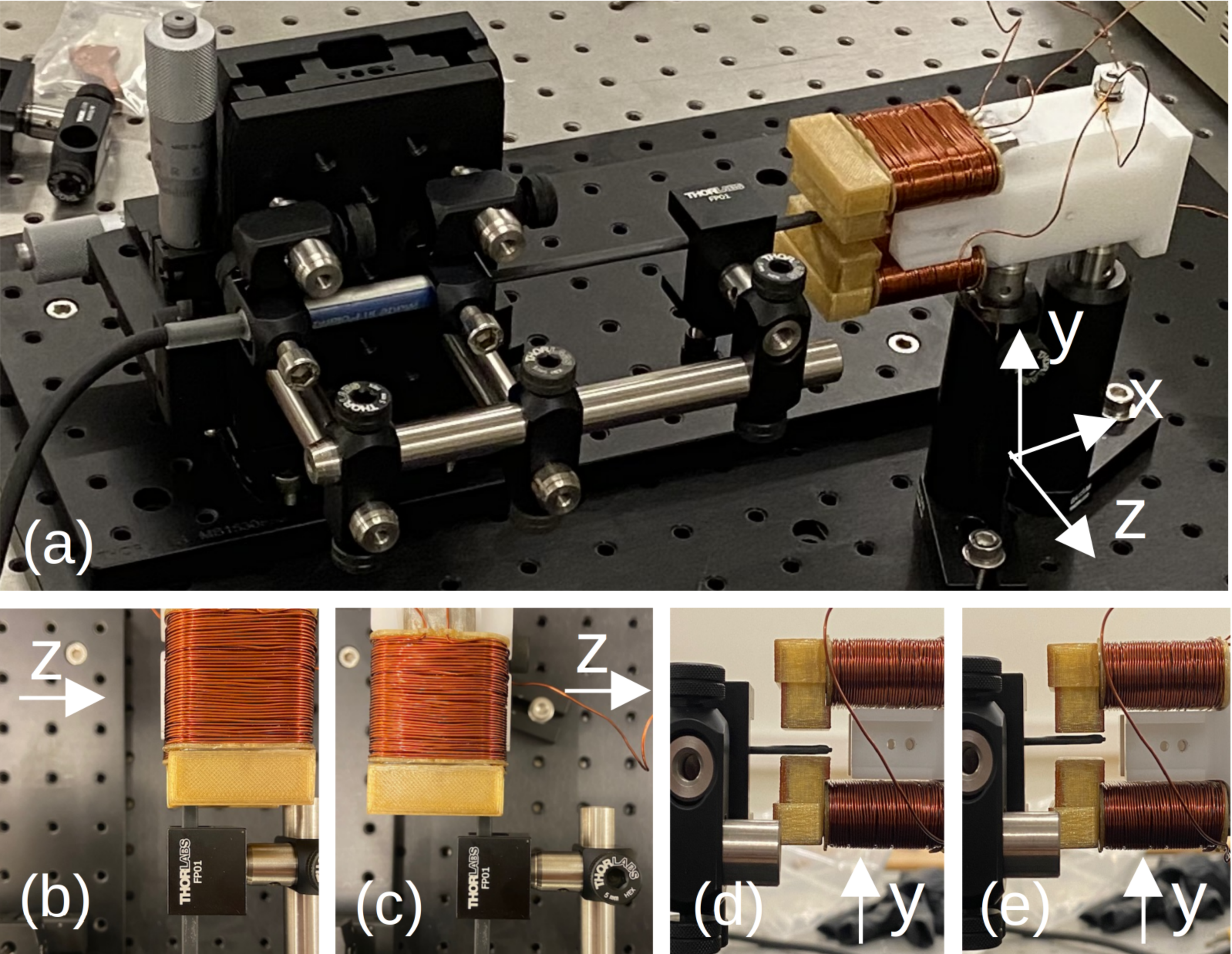}
\caption{Set-up used at the University of Oslo for external B-field characterisation: (a) set-up with magnetic probe, translation stage and electromagnet, (b--c) two extreme scan positions in $z$, (d--e) two extreme scan positions in $y$ ($xz$-planes).}\label{fig:adli_photo}
\end{figure}

The simulation-generated values given above have been cross-checked using a set-up at the University of Oslo, presented in Fig.~\ref{fig:adli_photo}.
The B-field is measured using a Hall-probe (ASONIK type SMS-102, radial), with precision of \SI{0.1}{mT}. It is mounted on a 3-axis translation stage, with \SI{1}{\micro\meter} precision. Each solenoid consists of 2 winding layers, ensuring a total winding density in each solenoid of $n_{\mathrm{sol}} = 2400$~turns/m. The solenoids are connected in series and the current is set on the DC-power source to $I_{\mathrm{sol}} = 1.5 \pm \SI{0.01}{A}$. The field generated by each solenoid in vacuum is calculated from the formula $B_{\mathrm{sol,vac}} = \mu{}_0 n_{\mathrm{sol}} I_{\mathrm{sol}} = \SI{4.45}{mT}$. Measurements are performed between the poles, without capillary and without electrodes (should not deviate from the final set-up, since copper is not magnetic). The measurement space steps in $[x,y,z]$ are [0.1, 1,5]~\SI{}{mm}. Since the absolute position of the B-field sensor was not known exactly, we calibrated our experiment by identifying the opposite positions where the B-field started to drop equally, and took the center of these points as the set-up center.

\begin{figure}[t]
\centering
\includegraphics[width=1.0\linewidth]{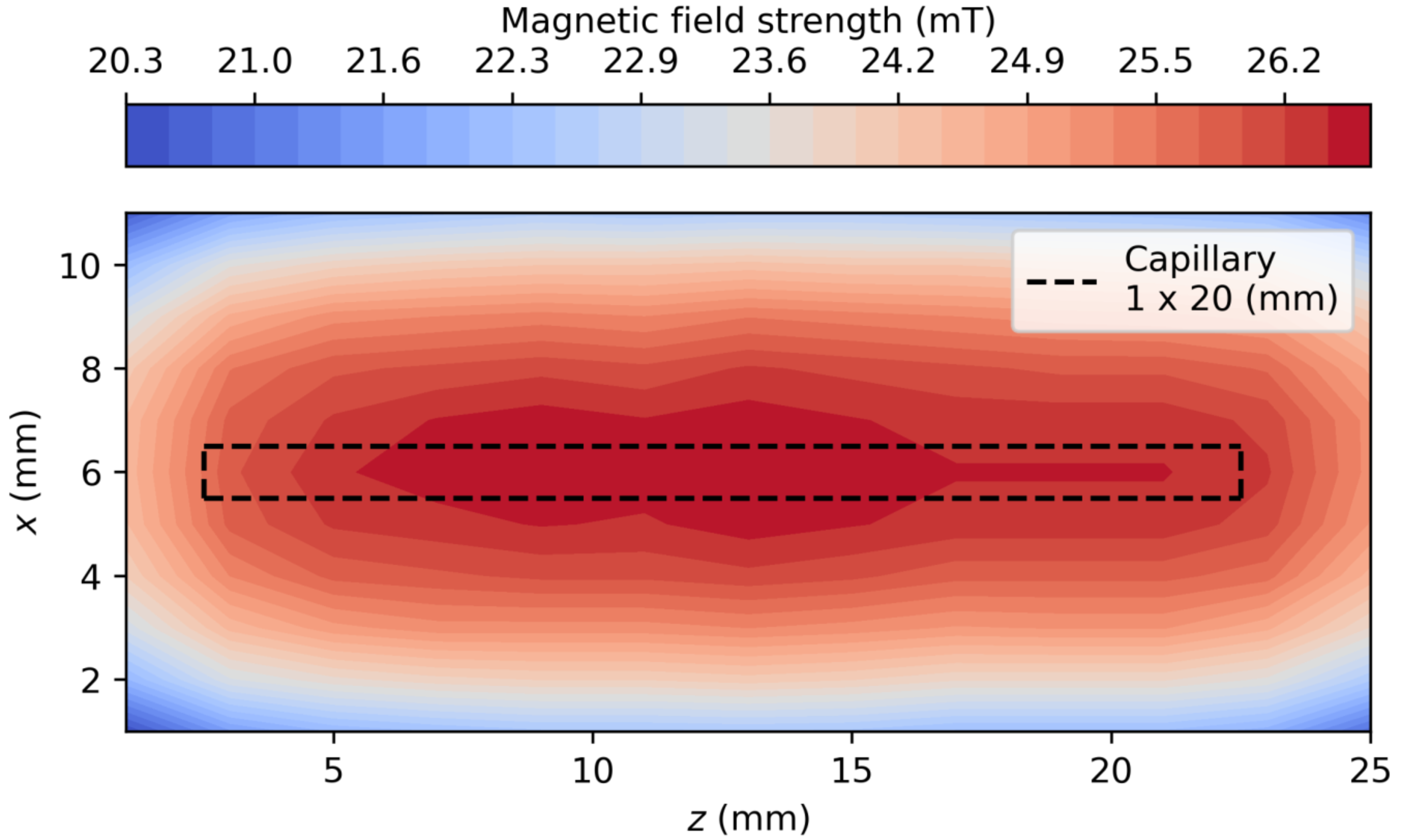}
\caption{Vertical B-field measurement at the plane in $y$ corresponding to the capillary center, performed without electrodes and without capillary. The virtual position of the capillary is added in dashed black.}\label{fig:adli_measurement}
\end{figure}

Measurements results are presented in Fig.~\ref{fig:adli_measurement}.
The vertical B-field has a mean value of \SI{26.4}{mT}, with variations up to 2\% within the capillary area defined in the figure.

Regarding the mean value itself, assuming a transfer from yoke arm to capillary of 23\% (from COMSOL simulations), this yields a field of \SI{115}{mT} in the arm. The relative permeability of the material then is evaluated to be roughly $\mu_r = 26$, which is consistent with regular steel. The design could provide an even higher B-field between the poles by using a higher-permeability material. One could even increase the current and the number of layers in the solenoids. 

Regarding the uniformity, the experimental measurement shows a satisfactory level of field uniformity in the plasma region, validating the magnetostatic COMSOL simulation to a few percent. Further simulations of plasma hydrodynamics and beam transport are however necessary to validate the effect of the provided level of external B-field uniformity.

\section{Experimental setup in the CLEAR User Facility at CERN}
\label{sec:setup}

In order to directly measure the B-field distribution generated by our nonlinear plasma lens, we use an external relativistic electron beam and measure its deflection by the lens. As already proposed and performed in~\cite{sjobak2021strong}, such a way to proceed allows to compute back the lens field. The angular deflection at the cell exit is measured as a transverse displacement on a downstream screen $(\Delta x,\Delta y)$, as a function of the lens offset $(x_l,y_l)$ with regards to the beam axis [with the point $(0,0)$ corresponding to the center of the lens coinciding with the beam axis]. Within the thin lens approximation, it can be expressed as follows \cite{lindstrom2018emittance}:
\begin{align}
\Delta{}x (x_l, y_l) &= + \frac{ecL\Delta{}s}{E} B_y(x_l, y_l)\label{eq:kick_x} \\
\Delta{}y(x_l, y_l) &= - \frac{ecL\Delta{}s}{E} B_x(x_l, y_l) \label{eq:kick_y}
\end{align}
with $e$ the electron charge, $c$ the speed of light in vacuum, $L$ the cell length, $\Delta{}s$ the distance from the capillary center to the screen and $E$ the beam energy (assuming ultra-relativistic beam particles).

The first step presented in this subsection \ref{ssec:commissioning} ensures that the design can be integrated in a beam-line, i.e., is vacuum-compatible and does not spark outside of the capillary. Subsection \ref{ssec:characterisation} describes the measurement of $B$ (sum of external field $B_{\mathrm{ext}}$ and pure discharge field $B_{\mathrm{dis}}$).

\subsection{Commissioning of the plasma lens}
\label{ssec:commissioning}

\begin{figure}[h]
\centering
\includegraphics[width=1.0\linewidth]{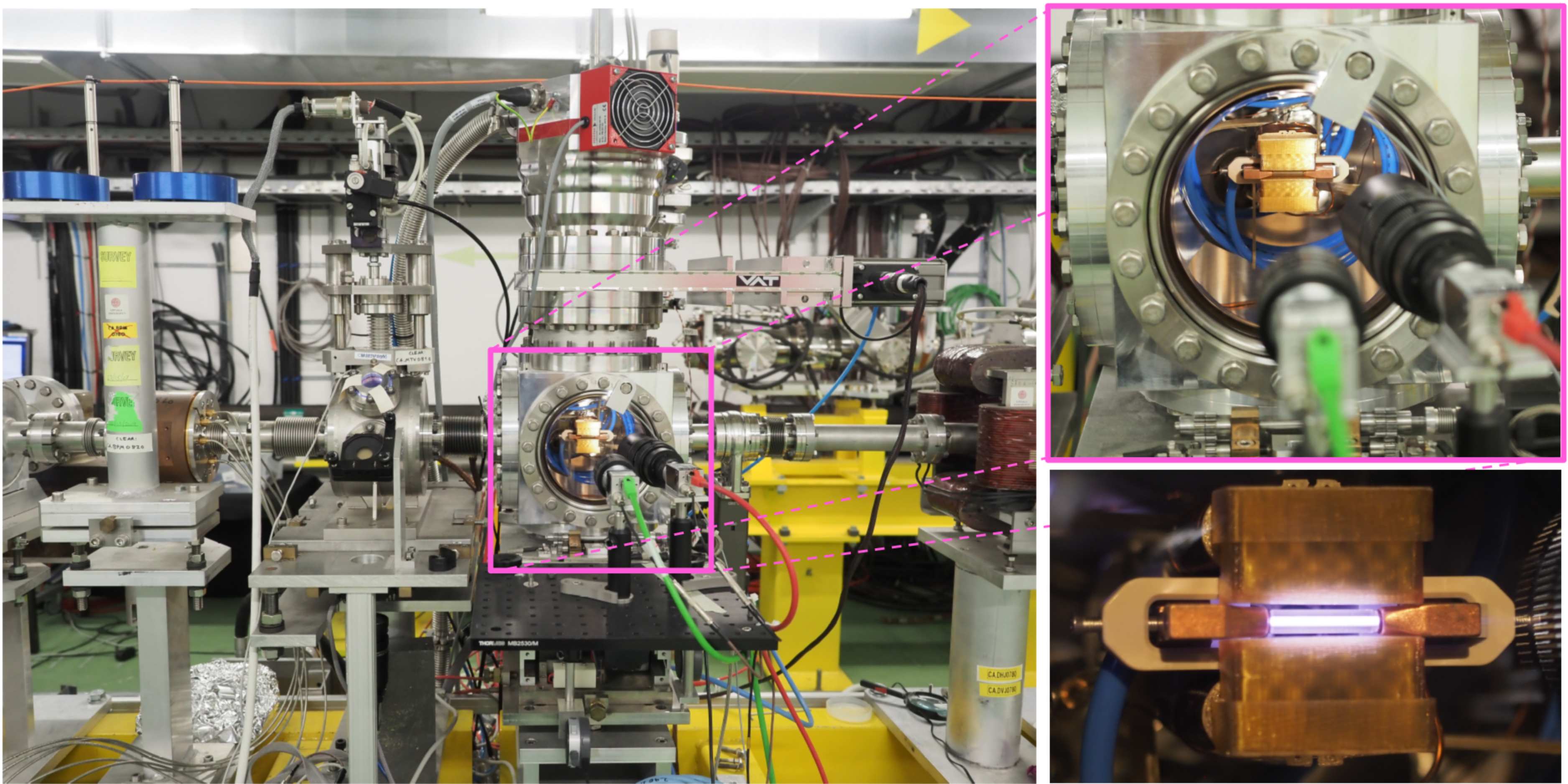}
\caption{Pictures of the nonlinear plasma lens integrated in the CERN CLEAR beam-line in September 2024. The electron beam goes from right to left.}\label{fig:lens_clear}
\end{figure}

We are performing the measurement at CLEAR at CERN \cite{Gamba2018,sjobak2019status}, offering electron with the following characteristics: 60--\SI{200}{MeV} energy, 0.01--\SI{50}{nC} charge, 1--100 bunches per pulse, 1--10 pulses per second, \SI{1}{ps}--\SI{50}{ns} pulse length, and a focus down to $50 \times \SI{50}{\micro\meter}$ at the plasma-lens entrance. The lens integrated in the beam-line is shown in Fig.~\ref{fig:lens_clear}. 

The commissioning has proven the validity of the design in terms of gas injection and sparking. Tested with argon, the pressure in the upstream gas buffer is set to approx.~\SI{20}{mbar}. Assuming 70\% pressure loss, as measured previously \cite{lindstrom2018emittance}, this implies \SI{\sim6}{mbar} in the capillary. The pressure in the vacuum chamber remains low enough for the secondary turbopump to remain at full speed. No discharge outside of the capillary was observed during operation. The discharge, around \SI{1}{kA}, shows a time-jitter below \SI{5}{ns}.

\subsection{Planning the plasma-lens characterisation}
\label{ssec:characterisation}

The second step of the experiment is the lens characterisation, i.e., measurement of the total field $B$ (sum of $B_{\mathrm{ext}}$ from the electromagnet and $B_{\mathrm{dis}}$ from the discharge). 

\begin{figure}[h]
\centering
\includegraphics[width=0.8\linewidth]{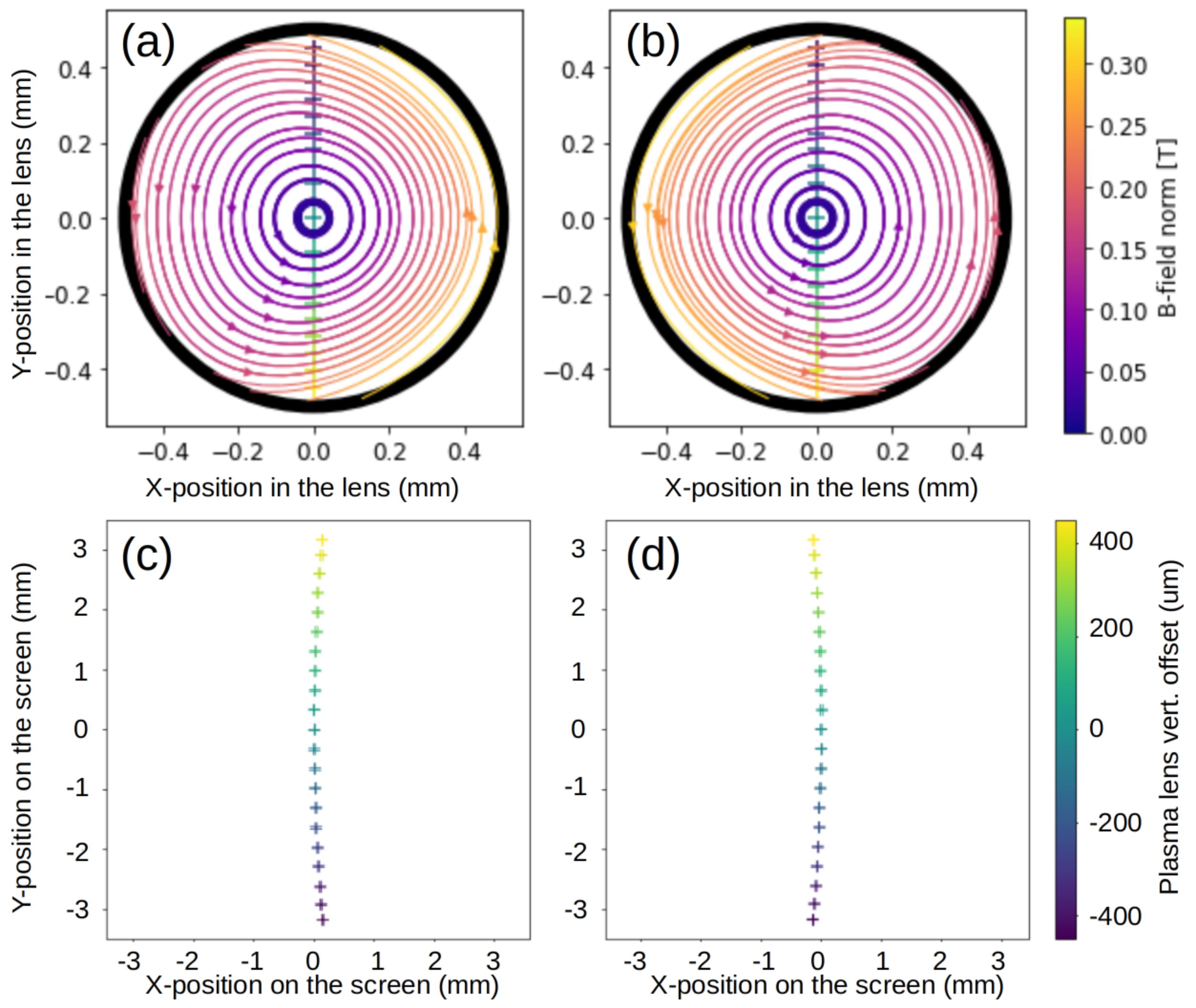}
\caption{Simulated results emulating an experiment with a $1\,$mm diameter capillary with: (a--b) theoretical B-field distribution in the lens with accentuated $1/D_x $ set to \SI{10}{\percent\per\mm} instead of \SI{1}{\percent\per\mm} for better visualisation of the nonlinearity, (c--d) predicted observation from tracking simulations on the screen downstream of the lens for an actual nonlinear factor $1/D_x = \SI{10}{\percent\per\mm}$. The points scanned are displayed as crosses in all sub-graphs. The effect of the external field $B_{\mathrm{ext}}$ from the electromagnet is neglected, resulting in a magnetic center coinciding with the cell geometrical center.}\label{fig:banana}
\end{figure}

Among the possible scans in the $xy$-plane, the most qualitatively visible outcome appears for a vertical scan centered on the lens axis. Here we approximate the geometrical axis to be also the magnetic axis, i.e., we neglect the constant axis-shift induced by the electromagnet through $B_{\mathrm{ext}}$. The scan positions are $(0,y_l)$, with resulting positions on the screen downstream described as follows (from Eq.~\ref{eq:kick_banana_x} and \ref{eq:kick_banana_y}):
\begin{align}
\Delta{}x (0, y_l) &= \frac{ecL\Delta{}s}{2 E} g_0 \frac{y_l^2}{D_x} \label{eq:kick_banana_x} \\
\Delta{}y(0, y_l) &= \frac{ecL\Delta{}s}{E} g_0 y_l \label{eq:kick_banana_y}.
\end{align}
Rough estimates using numbers from CLEAR, operating at maximum vertical offset, gives horizontal offsets on the screen of approximately 50--\SI{100}{\micro\m} (while the vertical offset on the screen is 2--\SI{3}{mm}). Tracking simulations were performed and the simulated results of such a scan are presented in Fig.~\ref{fig:banana}. A clear ``banana"-shape appears, which will be a key indicator of the nonlinearity.

The full characterisation of the lens requires a ``grid"-scan of the entire $xy$-plane. As a beginning, however, we focus on the demonstration of the nonlinearity.

\section{Conclusion and outlook}
\label{sec:conclusion}

In this article, we are proposing an active nonlinear plasma lens design, as part of a lattice for achromatic beam transport. The solution proposed to generate the nonlinearity is to use an external B-field, triggering Hall effect in the plasma and is supported by 1D hydrodynamics simulations in $H_2$. The design developed at the University of Oslo is presented and the external B-field generated by the electromagnet is simulated and experimentally measured. The characterisation of the full B-field distribution (i.e., the sum of the external field from the electromagnet and the magnetic field from the discharge current) inside the lens is planned to be conducted at CLEAR.

A preliminary experiment was already carried out in Sep.~2024 and results are under analysis. The objective at this stage is to experimentally observe and quantify a nonlinearity in focusing strength. In the medium-term, the lens should be used to focus a dispersed beam and prove its ability to match each energy component of the beam. Longer term, our lens design could experimentally validate the concept of achromatic transport when tested with the full lattice.

\section*{Acknowledgements}

The SPARTA project is funded by the European Union (ERC, SPARTA, 101116161). We acknowledge Sigma2 - the National Infrastructure for High-Performance Computing and Data Storage in Norway for awarding this project access to the LUMI supercomputer, owned by the EuroHPC Joint Undertaking, hosted by CSC (Finland) and the LUMI consortium. The work was also supported by the Research Council of Norway (NFR Grants No. 313770 and 310713).

We greatly thank:  the workshop I-Lab at the University of Oslo (J.~S. Ringnes, S.~R.~Solbak, H.~Borg) for their help in designing and manufacturing of the assembly and CERN for facilitating beam time at CLEAR (R.~Corsini, W.~Farabolini, A.~Gilardi, P.~Korysko, V.~Rieker).



\bibliographystyle{elsarticle-num} 
\bibliography{bibliography}

\end{document}